\documentclass[british,english]{article}
\usepackage[T1]{fontenc}
\usepackage[latin9]{inputenc}
\usepackage{geometry}
\usepackage{xcolor}
\usepackage{pdfcolmk}
\usepackage{array}
\usepackage{float}
\usepackage{multirow}
\usepackage{amsmath}
\usepackage{amssymb}
\usepackage{graphicx}
\PassOptionsToPackage{normalem}{ulem}
\usepackage{ulem}

\makeatletter

\providecommand{\tabularnewline}{\\}
\providecolor{lyxadded}{rgb}{0,0,1}
\providecolor{lyxdeleted}{rgb}{1,0,0}
\DeclareRobustCommand{\lyxsout}[1]{\ifx\\#1\else\sout{#1}\fi}

\@ifundefined{showcaptionsetup}{}{%
 \PassOptionsToPackage{caption=false}{subfig}}
\usepackage{subfig}
\makeatother

\usepackage{babel}
\begin{document}

\title{A bivariate likelihood approach for estimation of a pooled continuous
effect size from a heteroscedastic meta-analysis study}

\author{Osama Almalik\thanks{Department of Mathematics and Computer Science, Eindhoven University
of Technology. Postbus 513, 5600 MB Eindhoven, the Netherlands. E-mail:
O.Almalik@tue.nl. Tel.nr: +31625120134} \ and Edwin R. van den Heuvel\textsuperscript{{*}}\thanks{Preventive Medicine and Epidemiology, Department of Medicine, Boston
University, USA.}}
\maketitle
\begin{abstract}
\noindent The DerSimonian-Laird (DL) weighted average method has been
widely used for estimation of a pooled effect size from an aggregated
data meta-analysis study. It is mainly criticized for its underestimation
of the standard error of the pooled effect size in the presence of
heterogeneous study effect sizes. The uncertainty in the estimation
of the between-study variance is not accounted for in the calculation
of this standard error. Due to this negative property, many alternative
estimation approaches have been proposed in literature. One approach
was developed by Hardy and Thompson (HT), who implemented a profile
likelihood approach instead of the moment-based approach of DL. Others
have further extended the likelihood approach and proposed higher-order
likelihood inferences (e.g., Bartlett-type corrections). Likelihood-based
methods better address the uncertainty in estimating the between-study
variance than the DL method, but all these methods assume that the
within-study standard deviations are known and equal to the observed
standard error of the study effect sizes. Here we treat the observed
standard errors as estimators for the within-study variability and
we propose a bivariate likelihood approach that jointly estimates
the pooled effect size, the between-study variance, and the potentially
heteroscedastic within-study variances. We study the performance of
the proposed method by means of simulation, and compare it to DL,
HT, and the higher-order likelihood methods. Our proposed approach
appears to be less sensitive to the number of studies, and less biased
in case of heteroscedasticity.
\end{abstract}

\section{Introduction}

The DerSimonian-Laird (DL) method (DerSimonian and Laird, 1986) has
been and still is widely used to estimate a pooled effect size from
aggregated data meta-analysis studies. The method is a weighted average
of the study effect sizes, where the weights are the inverse study
variances (including both the within and between study variability).
The between-study variance component is estimated with a moment estimator.
The DL method was shown to be negatively biased when the number of
studies is small (Malzahn et al., 2000) and it does not account for
the uncertainty in estimating the between study variability (Hardy
and Thompson, 1996), potentially leading to liberal confidence intervals
for the pooled effect size (Veroniki et al. 2019).

Alternative methods have been proposed in literature to improve the
DL method (Viechtbauer, 2005; Veroniki et al., 2019). The most familiar
approach is the profile likelihood approach of Hardy and Thompson
(HT) (1996), where the study effect sizes are assumed normally distributed
and potentially heterogeneous, but with known within study variances.
The pooled effect size and the between-study variance component are
then estimated jointly. The authors constructed a confidence interval
for the pooled effect size that is based on the chi-square distribution
of a likelihood ratio statistic (Hardy and Thompson, 1996). It has
been shown that this profile likelihood approach has a closer to nominal
coverage probability than the DL method (Veroniki et al., 2019; Tanizaki,
2004).

However, the likelihood ratio statistic is only asymptotically chi-squared
distributed, and for small sample sizes the approximation might be
poor (Barndorff-Nielsen and Hall, 1988). For this reason, Noma (2011)
proposed a Bartlett-type correction for the likelihood ratio statistic
(Norma, 2011). \foreignlanguage{british}{Additionally, the author
proposed constructing confidence limits using the efficient score
statistic and a Bartlett-corrected efficient score function (Cox and
Hinkley, 1974; Guolo, 2012). These three methods for confidence intervals
of the pooled effect size showed conservative coverage probabilities,
especially when the number of studies is small, while the DL and the
HT methods had liberal coverage probabilities (Cox and Hinkley, 1974).}

\selectlanguage{british}%
The Bartlett-type correction of the likelihood ratio statistic is
only appropriate for exponential families (Guolo, 2012). The commonly
assumed random effects meta-analysis model is a member of the exponential
family in the unlikely case of equal within-study variances (Guolo,
2012). Guolo (2012) therefore applied an approximation to the Bartlett-type
correction introduced in Skovgaard (2001). This Guolo-Skovgaard (GS)
approximation produced conservative coverage probabilities in case
of a small number of studies, but its performance improved when the
number of studies increases (Guolo, 2012). In one comparative study,
the Bartlett-type correction method and the GS correction method were
found to produce similar results (Veroniki, 2019).

All of the methods discussed so far, assume that the within-study
standard deviation is given by the observed standard error of the
study effect size, while the true within-study variability is unknown
in practice. We will assume that the observed standard error is an
estimator of the true within-study variability having a chi-square
distribution function. We will introduce a bivariate likelihood approach
for estimation of the pooled effect size, the between-study variance,
and the within-study variances for heteroscedastic continuous outcomes.
Using a case study and a simulation study, we compare our method to
DL, HT, the Bartlett-type correction method, and the GS correction
method.

\selectlanguage{english}%
In section 2 we describe the\foreignlanguage{british}{ different approaches
from literature and }our proposed bivariate likelihood approach. The
approaches are illustrated on a real case study that was published
in literature before. Section 3 describes the simulation model we
used. It simulates meta-analysis studies with both heterogeneous effect
sizes and heteroscedastic standard errors. We believe that heteroscedastic
errors are common in practice, but are seldom simulated (van den Heuvel
et al., 2020). The results of the simulation study are provided in
Section 4 and a discussion is provided in Section 5.

\section{Statistical methods\label{sec:Statistical-methods}}

An aggregated data meta-analysis usually consists of\foreignlanguage{british}{
a set of $m$ effect sizes (e.g., mean differences, odds ratios, correlation
coefficients), accompanied with their standard errors and the degrees
of freedom (Cochran, 1954), i.e., we observe triplet $(Y_{i},S_{i},df_{i})$
for study $i=1,2,...,m$. It is typically assumed that the study effect
size $Y_{i}$ is distributed according to the meta-analysis model}
\begin{equation}
Y_{i}=\theta+U_{i}+\varepsilon_{i},\label{eq:random effects meta analysis model}
\end{equation}

with $\theta$ the true or pooled effect size, $U_{i}\sim N(0,\tau^{2})$
a random effect that is making the study effect sizes heterogeneous,
$\varepsilon_{i}\sim N(0,\sigma_{i}^{2})$ a residual, and all random
effects mutually independently distributed. The $\tau^{2}$ is the
variance component for the between-study variability and $\sigma_{i}^{2}$
is the variance component of the within-study variability. In literature
it is commonly assumed that the within-study variability $\sigma_{i}^{2}$
is known and given by $S_{i}^{2}$, but we believe that $S_{i}^{2}$
is at best an estimator of $\sigma_{i}^{2}$. We will assume that
$Y_{i}$ follows model (\ref{eq:random effects meta analysis model})
and the distribution of $df_{i}S_{i}^{2}/\sigma_{i}^{2}$ is approximately
chi-square with $df_{i}$ degrees of freedom. These assumptions typically
hold true when the study effect size is represented by a mean difference
with underlying normally distributed data (van den Heuvel et al.,
2020).

\selectlanguage{british}%
In sections \ref{subsec:DSL}, \ref{subsec:Existing-likelihood},
and \ref{subsec:Bivariate-distribution} we describe\foreignlanguage{english}{
the DL method, existing likelihood-based methods, and our bivariate
method }for estimating the effect size $\theta$ and constructing
95\% confidence intervals. Section \ref{subsec:Case-study} presents
a case study from literature where all methods are being demonstrated.
\selectlanguage{english}%

\subsection{The DerSimonian-Laird method\label{subsec:DSL}}

The DL method first estimates the between-study variance component
$\tau^{2}$ with the moment estimator given by
\[
\hat{\tau}_{DL}^{2}=\max\left[0,\frac{Q-\left(m-1\right)}{\sum_{i=1}^{m}w_{i}-\sum_{i=1}^{m}w_{i}^{2}/\sum_{i=1}^{m}w_{i}}\right],
\]

where $w_{i}=1/S_{i}^{2}$, $Q$ is Cochran's Q-statistic given by
$Q=\sum_{i=1}^{m}[(Y_{i}-\bar{Y})^{2}/S_{i}^{2}]$ (DerSimonian and
Laird 1986), and $\bar{Y}$ is the weighted average given by $\bar{Y}=\sum_{i=1}^{m}(Y_{i}/S_{i}^{2})/\sum_{i=1}^{m}(1/S_{i}^{2})$.
Then the pooled estimator $\hat{\theta}_{DL}$ of the effect size
$\theta$ is calculated using the estimator $\hat{\tau}_{DL}^{2}$.
The DL estimator is given by
\[
\hat{\theta}_{DL}=\left[\sum_{i=1}^{m}Y_{i}(\hat{\tau}_{DL}^{2}+S_{i}^{2})^{-1}\right]/\left[\sum_{i=1}^{m}(\hat{\tau}_{DL}^{2}+S_{i}^{2})^{-1}\right].
\]
A $\left(1-\alpha\right)\times100$ confidence interval on $\theta$
may be determined by $\hat{\theta}_{DL}\pm t_{\alpha/2,m-1}S_{DL}$,
with $t_{q,d}^{-1}$ the $q$\textsuperscript{th} upper quantile
of the $t$-distribution with $d$ degrees of freedom and $S_{DL}^{2}=[\sum_{i=1}^{m}1/(\hat{\tau}_{DL}^{2}+S_{i}^{2})]^{-1}$
the estimated variance of the pooled estimator $\hat{\theta}_{DL}$
having $m-1$ degrees of freedom. Note that it has been more common
in literature to use the normal quantile instead of the quantile of
the $t$-distribution (Brockwell and Gordon, 2007; Thorlund et al.,
2011; Jackson et al., 2010), but we believe that DerSimonian and Laird
were not explicit on this topic (DerSimonian and Laird, 1986) and
therefore did not rule out our preferred choice. We believe that our
choice is in line with the work of Cochran (Cochran, 1954), who proposed
to use the $t$-distribution with $m-1$ degrees of freedom instead
of the normal distribution, in particular in the presence of heterogeneity
(see also Mzolo et al. (2013)). The use of this $t$-distribution
is common when the corrected standard error of Hartung-Knapp-Sidik-Jonkman
is used (Sidik and Jonkman, 2005) . The standard error $S_{DL}$ is
then multiplied with a data-driven scaling factor $[\sum_{i=1}^{m}(Y_{i}-\hat{\theta}_{DL})^{2}/((\hat{\tau}_{DL}^{2}+S_{i}^{2})(n-1))]^{1/2}$
(see Sidik and Jonkman (2005). We will not study this corrected standard
error, even though it is often proposed as the preferred method In
't Hout et al., 2014), because the use of this corrected standard
error is not without criticism (Jackson et al., 2017; Partlett and
Riley, 2017). Furthermore, comparisons between the use of this corrected
standard error and the traditional DerSimonian-Laird method typically
studied normal quantiles and never investigated the influence of the
proposed $t$-distribution alone.

To obtain the estimates $\hat{\tau}_{DL}^{2}$ and $\hat{\theta}_{DL}$
and the confidence limits on $\theta$ from data in our simulation
study, we programmed the method in SAS, since most {[}R{]} packages
seem to have incorporated a normal quantile or otherwise use the corrected
standard error with the $t$-distribution (e.g., ``meta'' (Schwarzer,
2007) and ``metafor'' (Viechtbauer, 2010).

\subsection{Existing likelihood-based methods\label{subsec:Existing-likelihood}}

Three likelihood based approaches for parameter estimation and confidence
intervals have been proposed in literature. They all make use of the
same maximum likelihood estimators for the parameters $\theta$ and
$\tau^{2}$, which is based on the procedure of Hardy and Thompson
(1996), but they differ in the construction of confidence intervals.

\subsubsection{The Hardy-Thompson method\label{subsec:Hardy-Thompson}}

The log-likelihood function that was proposed in Hardy and Thompson
(1996) is given by
\begin{equation}
l(\theta,\tau^{2})=-\frac{1}{2}m\log(2\pi)-\frac{1}{2}\sum_{i=1}^{m}\log(\tau^{2}+S_{i}^{2})-\frac{1}{2}\sum_{i=1}^{m}(Y_{i}-\theta)^{2}/(\tau^{2}+S_{i}^{2}).\label{eq:Hardy and Thompson}
\end{equation}
It shows that the within-study variances $\sigma_{i}^{2}$ are assumed
known and equal to $S_{i}^{2}$. Maximizing (\ref{eq:Hardy and Thompson})
with respect to $\theta$ and $\tau^{2}$ results in solving the following
two equations iteratively:
\begin{equation}
\begin{array}{rcl}
\theta & = & \left[\sum_{i=1}^{m}Y_{i}(\tau^{2}+S_{i}^{2})^{-1}\right]/\left[\sum_{i=1}^{m}(\tau^{2}+S_{i}^{2})^{-1}\right],\\
\tau^{2} & = & \left[\sum_{i=1}^{m}((Y_{i}-\theta)^{2}-S_{i}^{2})(\tau^{2}+S_{i}^{2})^{-2}\right]/\left[\sum_{i=1}^{m}(\tau^{2}+S_{i}^{2})^{-2}\right].
\end{array}\label{eq:likelihood-equations}
\end{equation}
The two solutions are Hardy and Thompson's (HT) maximum likelihood
estimators $\hat{\theta}_{HT}$ and $\hat{\tau}_{HT}^{2}$. For the
construction of confidence regions on $(\theta,\tau^{2})$ a kind
of log-likelihood ratio statistic $T_{HT}(\theta,\tau^{2})$ was proposed:
\begin{equation}
T_{HT}(\theta,\tau^{2})=-2[l(\theta,\tau^{2})-l(\hat{\theta}_{HT},\hat{\tau}_{HT}^{2})].\label{eq:LRT statistic}
\end{equation}
It is assumed that $T_{HT}(\theta,\tau^{2})$ is chi-square distributed
with $2$ degrees of freedom. All pairs of values $(\theta,\tau^{2})$
that would satisfy $T_{HT}(\theta,\tau^{2})<\chi_{2}^{-2}(1-\alpha)$,
with $\tau^{2}\geq0$ and $\chi_{d}^{-2}(q)$ the $q$\textsuperscript{th}
upper quantile of the chi-square distribution with $d$ degrees of
freedom, form the $100\%\times(1-\alpha)$ confidence region on $(\theta,\tau^{2})$
(Hardy and Thompson, 1996) .

To obtain confidence intervals for $\theta$ and $\tau^{2}$ separately,
a profile likelihood function was considered. Here we focus on the
$100\%\times(1-\alpha)$ confidence interval for $\theta$, but a
similar approach can be applied to $\tau$. If we assume that $\theta$
is given, we could maximize the log-likelihood function in (\ref{eq:Hardy and Thompson})
for $\tau$ first, resulting in the constraint maximum likelihood
estimator $\hat{\tau}^{2}(\theta)$. Substituting this estimator in
(\ref{eq:Hardy and Thompson}) results in the profile log-likelihood
function $\tilde{l}(\theta)\equiv l(\theta,\hat{\tau}^{2}(\theta))$.
The profile log-likelihood ratio statistic for $\theta$ is then defined
as
\begin{equation}
\tilde{T}_{HT}(\theta)=-2[\tilde{l}(\theta)-\tilde{l}(\hat{\theta}_{HT})].\label{eq:profile LRT}
\end{equation}
All values of $\theta$ that would satisfy inequality $\tilde{T}_{HT}(\theta)<\chi_{1}^{-2}(1-\alpha)$
would form the $100\%\times(1-\alpha)$ confidence interval for $\theta$.

The {[}R{]} package ``metaplus'' (Beath, 2016) can determine the
maximum likelihood estimators $\hat{\theta}_{HT}$ and $\hat{\tau}_{HT}^{2}$,
the confidence region for $(\theta,\tau^{2})$, and the two confidence
intervals for $\theta$ and $\tau^{2}$ from real data. We have used
this for the case study and simulation study.

\subsubsection{The Noma-Bartlett method}

The profile likelihood approach for $\theta$ mentioned in Section
\ref{subsec:Hardy-Thompson} is considered a first-order likelihood
inference method (Guolo, 2012). Higher-order asymptotic methods for
the proposed profile likelihood ratio statistic will provide more
accurate inference (Barndorff-Nielsen and Hall, 1988; Cox and Hinkley,
1974), in particular for smaller values of $m$. Norma (2011) applied
a Bartlett-type correction (Barndorff-Nielsen and Hall, 1988) to the
profile likelihood ratio statistic $\tilde{T}_{HT}(\theta)$ in (\ref{eq:profile LRT})
by normalizing it with a constant that depends on the constraint maximum
likelihood estimator $\hat{\tau}^{2}(\theta)$. The Noma-Bartlett
(NB) method uses this corrected likelihood ratio statistic, which
is given by $\tilde{T}_{NB}(\theta)=\tilde{T}_{HT}(\theta)/[1+2C(\hat{\tau}^{2}(\theta))]$,
with
\[
C(\tau^{2})=\left[\sum_{i=1}^{m}(S_{i}^{2}+\tau^{2})^{-3}\right]/\left[\sum_{i=1}^{m}(S_{i}^{2}+\tau^{2})^{-1}\sum_{i=1}^{m}(S_{i}^{2}+\tau^{2})^{-2}\right].
\]

The $100\%\times(1-\alpha)$ confidence interval for $\theta$ is
formed by all $\theta$'s satisfying $\tilde{T}_{NB}(\theta)<\chi_{1}^{-2}(1-\alpha)$.
For the case study and our simulation study we obtained the estimates
of the pooled effect size $\theta$ and the NB confidence interval
with the {[}R{]} package ``pimeta'' (Nagashima et al., 2019). Note
that the NB method uses the estimators $\hat{\theta}_{HT}$ and $\hat{\tau}_{HT}^{2}$
of Hardy and Thompson, but provides only an alternative confidence
interval for $\theta$.

\subsubsection{The Guolo-Skovgaard method}

Instead of using the profile likelihood ratio statistic $\tilde{T}_{HT}(\theta)$
in (\ref{eq:profile LRT}), a signed profile likelihood ratio statistic
can be used:
\begin{equation}
\tilde{r}_{G}(\theta)=\mathrm{sign}(\hat{\theta}_{HT}-\theta)\sqrt{l(\hat{\theta}_{HT},\hat{\tau}_{HT}^{2})-l(\theta,\hat{\tau}^{2}(\theta))}.\label{eq:Signed Likelihood ratio statistic}
\end{equation}
The statistic $\tilde{r}_{G}(\theta)$ is approximately normally distributed
(Guolo, 2012). Thus the set of values $\theta$ for which inequalities
$z_{\alpha/2}\leq\tilde{r}_{G}(\theta)\leq z_{1-\alpha/2}$ hold true,
with $z_{q}$ the $q$\textsuperscript{th} quantile of a standard
normal distribution, provides a $100\%\times(1-\alpha)$ confidence
interval for $\theta$.

Alternatively, a Skovgaard correction to the signed profile likelihood
ratio statistic in (\ref{eq:Signed Likelihood ratio statistic}) can
be applied in a random-effects meta-analysis. This Guolo-Skovgaard
corrected statistic is given by
\[
\tilde{r}_{GS}(\theta)=\tilde{r}_{G}(\theta)+[\tilde{r}_{G}(\theta)]^{-1}\log(\tilde{u}(\theta)/\tilde{r}_{G}(\theta)),
\]

with $\tilde{u}(\theta)=\left[S^{-1}(\theta)q(\theta)\right]_{1}|I(\hat{\theta}_{HT},\hat{\tau}_{HT}^{2})|^{1/2}|J(\hat{\theta}_{HT},\hat{\tau}_{HT}^{2})|^{-1}|S(\theta)||I_{22}(\theta,\hat{\tau}^{2}(\theta))|^{-1/2}$,
$S(\theta)$ the $2\times2$ matrix given by
\[
S(\theta)=\left(\begin{array}{cc}
\sum_{i=1}^{m}(S_{i}^{2}+\hat{\tau}^{2}(\theta))^{-1} & \sum_{i=1}^{m}(\hat{\theta}_{HT}-\theta)(S_{i}^{2}+\hat{\tau}^{2}(\theta))^{-2}\\
0 & \left[\sum_{i=1}^{m}(S_{i}^{2}+\hat{\tau}^{2}(\theta))^{-2}\right]/2
\end{array}\right),
\]

$q(\theta)$ the vector given by
\[
q(\theta)=\left(\begin{array}{c}
\sum_{i=1}^{m}(\hat{\theta}_{HT}-\theta)(S_{i}^{2}+\hat{\tau}^{2}(\theta))^{-1}\\
-\sum_{i=1}^{m}\left[(S_{i}^{2}+\hat{\tau}_{HT}^{2})^{-1}-(S_{i}^{2}+\hat{\tau}^{2}(\theta))^{-1}\right]/2
\end{array}\right),
\]

$I(\theta,\tau^{2})$ the $2\times2$ Fisher information matrix, $I_{22}(\theta,\tau^{2})$
the second diagonal element of $I(\theta,\tau^{2})$, $J(\theta,\tau^{2})$
the Hessian matrix (i.e., $I(\theta,\tau^{2})=\mathbb{-E}J(\theta,\tau^{2})$),
and $\left[S^{-1}(\theta)q(\theta)\right]_{1}$ the first element
of the vector $S^{-1}(\theta)q(\theta)$. The Guolo-Skovgaard (GS)
$100\%\times(1-\alpha)$ confidence interval for $\theta$ is obtained
by the set of values of $\theta$ that satisfies $z_{\alpha/2}\leq\tilde{r}_{GS}(\theta)\leq z_{1-\alpha/2}$.
These confidence limits will be calculated from data using {[}R{]}
package ``metaLik'' (Guolo and Varin, 2012). Also the GS method
uses the estimators $\hat{\theta}_{HT}$ and $\hat{\tau}_{HT}^{2}$
of Hardy and Thompson, and constructs only an alternative confidence
interval for $\theta$.

\subsection{A bivariate distribution method\label{subsec:Bivariate-distribution}}

The methods discussed in Sections \ref{subsec:DSL} and \ref{subsec:Existing-likelihood}
provide estimators and confidence intervals for the parameters $\theta$
and $\tau^{2}$ conditionally on $\sigma_{i}^{2}=S_{i}^{2}$. We believe
that $S_{i}^{2}$ should be viewed as an estimator for $\sigma_{i}^{2}$.
In this view it will not be likely equal to $\sigma_{i}^{2}$. Treating
$S_{i}^{2}$ as an estimator for $\sigma_{i}^{2}$, instead of conducting
a conditional analysis, has been acknowledged in literature (Hardy
and Thompson, 1996), but it has also been refuted, since it would
not or marginally affect the calculation of confidence intervals for
$\theta$ compared to the conditional analysis (Hardy and Thompson,
1996). It is argued that the estimator for the between-study variance
plays a more dominant role in the calculation of confidence intervals
on $\theta$ than the within-study variances. Nevertheless, we propose
a bivariate distribution method in which both $Y_{i}$ and $S_{i}^{2}$
are jointly modelled.

We assume that $Y_{i}$ follows model (\ref{eq:random effects meta analysis model})
and $S_{i}^{2}$ has approximately a chi-square distribution, i.e.,
$df_{i}S_{i}^{2}/\sigma_{i}^{2}\sim\chi_{df_{i}}^{2}$, with $df_{i}$
the degrees of freedom for $S_{i}^{2}$. We assume that $df_{i}$
is either observed or can be calculated from the aggregated information.
Furthermore, we assume that $\sigma_{i}^{2}\approx\sigma^{2}\eta_{i}$,
with $\eta_{i}>0$ a known value that would typically depend on the
sample size of study $i$. Thus we assume that the residual variances
in (\ref{eq:random effects meta analysis model}) are considered heteroscedastic
across studies as a consequence of different study sizes, but they
share a common within-study variance parameter $\sigma^{2}$. The
assumption $\sigma_{i}^{2}\approx\sigma^{2}\eta_{i}$ is in line with
literature on pooling estimates from biological assays (Cochran, 1954)
and helps us maintain a parsimonious model. Estimating $m$ variance
parameters $\sigma_{i}^{2}$ may result in an overfit and will lead
to numerical complexities.

One example is $\eta_{i}=[n_{i}-3]^{-1}$, with $n_{i}$ the total
sample size for study $i$. This choice fits with pooling Fisher's
$z$ transformed correlation coefficients. Estimation of the variance
parameter $\sigma^{2}$ is then expected to be close to one. An other
example is $\eta_{i}=[n_{i0}^{-1}+n_{i1}^{-1}]$, with $n_{ij}$ the
sample size for the binary exposure $j\in\{0,1\}$. This choice fits
with pooling mean differences. The variance parameter $\sigma^{2}$
represents the between-participant variation within studies (van den
Heuvel et al., 2020). More generally, we may always consider $\eta_{i}=[df_{i}]^{-1}$,
where $df_{i}$ is then viewed as the effective sample size of study
$i$. The variance parameter $\sigma^{2}$ would then become a \textit{nuisance
parameter} as a measure of within-study variability without having
a direct meaning to the underlying individual data from the studies.
Note that our simulation model will be more heteroscedastic than what
we assume in this estimation approach to verify the robustness of
our approach.

The log-likelihood function for the bivariate distribution of $(Y_{i},S_{i}^{2})$
is given by
\begin{equation}
\begin{array}{l}
l\left(\theta,\tau^{2},\sigma^{2}\right)\approx-\frac{1}{2}\left[m\log(2\pi)+\sum_{i=1}^{m}\left(\log(\tau^{2}+\sigma_{i}^{2})+(Y_{i}-\theta)^{2}/(\tau^{2}+\sigma_{i}^{2})\right)\right.\\
\qquad\qquad\qquad\qquad\qquad\left.+df\log(2)+\sum\left(2\log\varGamma(df_{i}/2)+(df_{i}-2)\log\left(\chi_{i}^{2}\right)+\chi_{i}^{2}\right)\right],
\end{array}\label{eq:Bivariate Likelihood function}
\end{equation}
with $\chi_{i}^{2}=df_{i}S_{i}^{2}/\sigma_{i}^{2}$ the chi-square
statistic, $df=\sum_{i=1}^{m}df_{i}$ the total number of degrees
of freedom, and $\varGamma$ the gamma function. Note that the sum
$\sum_{i=1}^{m}\chi_{i}^{2}$ in the likelihood (\ref{eq:Bivariate Likelihood function})
is also chi-square distributed with $df$ degrees of freedom (Moschopoulos,
1985). Calculating the likelihood equations for the estimation of
the parameters $\theta$, $\tau^{2}$, and $\sigma^{2}$, leads to
the two equations in (\ref{eq:likelihood-equations}) with $S_{i}^{2}$
replaced by $\sigma_{i}^{2}=\sigma^{2}/df_{i}$ and additionally to
a third equation
\[
\sum_{i=1}^{m}\dfrac{(Y_{i}-\theta)^{2}-(\tau^{2}+\sigma_{i}^{2})}{df_{i}(\tau^{2}+\sigma_{i}^{2})^{2}}=\sum_{i=1}^{m}\dfrac{(df_{i}-2)\sigma_{i}^{2}-df_{i}S_{i}^{2}}{df_{i}\sigma_{i}^{4}}.
\]
Here $\sigma^{2}$ can be obtained by applying the Newton-Raphson
method (Choi and Wette, 1969) if $\theta$ and $\tau^{2}$ would be
given. Estimation of all three parameters $\theta$, $\tau^{2}$,
and $\sigma^{2}$ can be obtained with the procedure NLMIXED of SAS
software. The ML estimators of our bivariate distribution method are
referred to as $\hat{\theta}_{BD}$, $\hat{\tau}_{BD}^{2}$, and $\hat{\sigma}_{BD}^{2}$.
The programming codes for procedure NLMIXED are provided in the appendix.

An asymptotic $100\%\times(1-\alpha)$ confidence interval on $\theta$
can also be provided by the SAS procedure NLMIXED and it is given
by $\hat{\theta}_{BD}\pm t_{m-1,\alpha/2}^{-1}\hat{SE}(\hat{\theta}_{BD})$,
with $t_{d,q}^{-1}$ the $q$\textsuperscript{th} upper quantile
of the t-distribution with $d$ degrees of freedom, and $\hat{SE}(\hat{\theta}_{BD})$
the estimated asymptotic standard error of the estimator $\hat{\theta}_{BD}$
(SAS Institute, 1996). SAS uses the number of random effects minus
one ($m-1$) as the default number of degrees of freedom.

We do realize that the proposed bivariate method requires more input
than the other described methods, since the number of degrees of freedom
$df_{i}$ associated with the within-study variance estimate $S_{i}^{2}$
is required in our approach. However, in practice we expect that meta-analysts
may have access to this information or otherwise can calculate or
create the appropriate degrees of freedom for $S_{i}^{2}$.

\subsection{Case study from literature\label{subsec:Case-study}}

To illustrate the approaches, we applied them to a meta-analysis on
mean platelet volume (MPV) and coronary artery disease (CAD) (Sansanayudh
et al., 2014). One of their aims was to conduct a systematic review
and meta-analysis comparing mean differences in MPV between patients
(CAD) and controls. Forty studies were included in this meta-analysis
based on the authors eligibility criteria, but 36 studies reported
a mean difference and only 31 studies compared the mean MPV between
CAD patients and controls. We used the data of these 31 studies (see
Figure 2 in Sansanayudh et al. (2014) ). For these studies we extracted
the means, standard deviations, and sample sizes for patients and
controls and calculated the mean difference, an estimate of the standard
error, and an accompanied degrees of freedom (see Section \ref{sec:Simulation-model}).

The pooled mean difference with their 95\% confidence intervals and
the estimate for the between study variance $\tau^{2}$ for our five
approaches are presented in Table \ref{tab:Example} The estimates
from HT, NB, and GS are all equal, since they are the maximum likelihood
estimates for likelihood function (\ref{eq:Hardy and Thompson}).
They only differ in the calculation of confidence intervals. DL has
the smallest pooled estimate, the smallest estimate for $\tau^{2}$,
and the narrowest 95\% confidence interval. The 95\% confidence intervals
for NB and GS are slightly wider than the 95\% confidence interval
of HT, which is the intention of these two methods. The pooled estimate
of BD is very close to the estimate of HT, but the 95\% confidence
interval is slightly wider than the 95\% confidence interval of NB
and GS. The estimate of the between-study variance of BD is also slightly
larger than the estimate of HT. The reported pooled estimate in (Sansanayudh
et al., 2014) based on DL was 0.70 (0.55; 0.85). The 95\% confidence
interval is smaller than ours, since we used a $t$-distribution instead
of the normal distribution.
\begin{center}
\begin{table}[h]
\caption{\label{tab:Example}Combined estimate of mean difference of MPV, along
with 95\% confidence limits and the between-study variance estimate}

\centering{}%
\begin{tabular}{|c|c|c|}
\hline
Method & $\hat{\theta}$ with 95\% confidence limits & $\hat{\tau}^{2}$\tabularnewline
\hline
\hline
DL & 0.6990 (0.5437; 0.8542) & 0.1532\tabularnewline
\hline
HT & 0.7031 (0.5259; 0.8835) & 0.2137\tabularnewline
\hline
NB & 0.7031 (0.5198; 0.8898) & 0.2137\tabularnewline
\hline
GS & 0.7031 (0.5216; 0.8901) & 0.2137\tabularnewline
\hline
BD & 0.7033 (0.5004; 0.9064) & 0.2284\tabularnewline
\hline
\end{tabular}
\end{table}
\par\end{center}

\section{Simulation model\label{sec:Simulation-model}}

We use a simulation model to generate data from an individual participant
data (IPD) meta-analysis. The IPD is used to calculate a study effect
size $Y_{i}$, an accompanied standard error $S_{i}$, and its associated
degrees of freedom $df_{i}$. The aggregated data is then pooled using
the methods described in Section \ref{sec:Statistical-methods}. Different
settings for the IPD model parameters were selected. A number of 1000
simulation runs were generated for each setting. For each simulation
run, the parameter $\theta$ is estimated and accompanied with a 95\%
confidence interval using the methods described in Section \ref{sec:Statistical-methods}.
We present the bias, mean squared error (MSE), and the coverage probability
for the main parameter $\theta$.

\subsection{Simulation model}

We simulated an IPD meta-analysis with $m$ studies. The sample size
$n_{i}$ for study $i=1,\cdots,m$ varied from study to study. This
sample size was drawn from an overdispersed Poisson distribution,
i.e., $n_{i}|\gamma_{i}\sim\mathrm{Poi}(\lambda\exp\{0.5\gamma_{i}\})$,
with $\gamma_{i}\sim\Gamma(a_{0},b_{0})$ drawn from a gamma distribution.
Then within each study the participants are randomly allocated to
two groups (e.g., treatments) with probabilities $p$ and $1-p$,
resulting in $n_{i0}$ participants in the control group (i.e., $n_{i0}|n_{i}\sim\mathrm{Bin}(n_{i},p$))
and $n_{i1}=n_{i}-n_{i0}$ participants in the exposed group. A continuous
response $Y_{ijk}$ for individual $k$ $(=1,\cdots,n_{ij})$, in
group $j$ $(=0,1)$, of study $i$ is then simulated according to
a heteroscedastic linear mixed effects model (Quintero and Lesaffre,
2017; Davidian and Carroll, 1987):
\begin{equation}
Y_{ijk}=\mu_{j}+U_{ij}+\xi_{j}\exp\left(V_{i}\right)\epsilon_{ijk},\label{eq:general model}
\end{equation}
with $\mu_{j}$ the mean of group $j$, $U_{ij}$ a study-specific
random effect for group $j$, $\xi_{j}^{2}$ a group-specific residual
variance parameter, $V_{i}$ a random effect for residual heteroscedasticity
across studies, and $\epsilon_{ijk}\sim N\left(0,1\right)$ standard
normally distributed and independent of random effects $U_{i0}$,
$U_{i1}$, and $V_{i}$. It is assumed that $(U_{i0},U_{i1},V_{i})^{T}$
has a multivariate normal distribution with means $0$ and variance-covariance
matrix $\varSigma$ given by
\[
\varSigma=\left(\begin{array}{ccc}
\sigma_{0}^{2} & \rho_{M}\sigma_{0}\sigma_{1} & \rho_{V}\sigma_{0}\sigma_{2}\\
\rho_{M}\sigma_{0}\sigma_{1} & \sigma_{1}^{2} & \rho_{V}\sigma_{1}\sigma_{2}\\
\rho_{V}\sigma_{0}\sigma_{2} & \rho_{V}\sigma_{1}\sigma_{2} & \sigma_{2}^{2}
\end{array}\right).
\]
The value of $\rho_{M}$ represents the correlation between the study-specific
random effects $U_{i0}$ and $U_{i1}$ for the exposed and the control
group, respectively. The value $\rho_{V}$ represents the correlation
between the study mean and the logarithm of the random heteroscedastic
residual variance.

There are two forms of residual heteroscedasticity in IPD model (\ref{eq:general model}).
One is at the level of the participant and introduced via parameter
$\xi_{j}^{2}$ and the other one is at the level of the study introduced
via the random term $\exp(V_{i})$. The variance $\xi_{j}^{2}$ indicates
a fixed heteroscedasticity in variability between individuals for
the two groups (i.e., the group affects both the level and the variability)
and is consistent across studies, while $\exp(V_{i})$ indicates a
random heteroscedasticity across studies and it is consistent within
studies (i.e., individuals are more or less alike within studies).

The observed study effect measure aggregated at the study level is
given by the raw mean difference $Y_{i}=\bar{Y}_{i0.}-\bar{Y}_{i1.}$
for study $i$, where $\bar{Y}_{ij.}=\sum_{k=1}^{n_{ij}}Y_{ijk}/n_{ij}$
is the average value for group $j$ in study $i$. Based on model
(\ref{eq:general model}), the observed study effect can be written
into the well-known random effects model\footnote{In the random effects model it is often assumed that the random variables
$U_{i}$ and $\varepsilon_{i}$ are independent and normally distributed,
but due to our random heteroscedastic variable $\exp\{V_{i}\}$ both
assumptions will be violated.} for meta-analysis studies (Brockwell and Gordon, 2007)
\begin{equation}
Y_{i}=\theta+U_{i}+\varepsilon_{i},\label{eq:RE-model-1}
\end{equation}
with $\theta=\mu_{0}-\mu_{1}$ the overall mean difference, $U_{i}\equiv U_{i0}-U_{i1}$
represents the study effect size heterogeneity, $\varepsilon_{i}=\exp(V_{i})(\xi_{0}\bar{\epsilon}_{i0.}-\xi_{1}\bar{\epsilon}_{i1.})$
is the within-study residual with $\bar{\epsilon}_{ij.}=\sum_{k=1}^{n_{ij}}\epsilon_{ijk}/n_{ij}$.
If $\rho_{M}=1$ and $\sigma_{0}=\sigma_{1}$, $U_{i0}-U_{i1}$ is
degenerate in zero or non-existent, while for all other settings of
$\rho_{M}<1$, $\sigma_{0}>0$, and $\sigma_{1}>0$ it will lead to
heterogeneous study effect sizes. Without the existence of $V_{i}$,
the residuals $\varepsilon_{i}$ in (\ref{eq:RE-model-1}) are still
heteroscedastic across studies ($\mathsf{VAR}(\varepsilon_{i})=\xi_{0}^{2}/n_{i0}+\xi_{1}^{2}/n_{i1}$),
unless sample sizes are consistent across studies.

The estimated standard error $S_{i}$ for the study effect size $Y_{i}$
is given by $S_{i}^{2}=S_{i0}^{2}/n_{i0}+S_{i1}^{2}/n_{i1}$, where
$S_{ij}^{2}=\sum_{k=1}^{n_{ij}}(Y_{ijk}-\bar{Y}_{ij.})^{2}/(n_{ij}-1)$
is the sample variance for group $j$ in study $i$. Here we allow
that the variability between individuals within treatments could be
different. The variance $S_{i}^{2}$ can be rewritten into
\begin{equation}
S_{i}^{2}=\exp(2V_{i})(\xi_{0}^{2}s_{i0}^{2}/n_{i0}+\xi_{1}^{2}s_{i1}^{2}/n_{i1}),\label{eq:SE}
\end{equation}
with $(n_{ij}-1)s_{ij}^{2}=\sum_{k=1}^{n_{ij}}(\epsilon_{ijk}-\bar{\epsilon}_{ij.})^{2}$
chi-square distributed with $n_{ij}-1$ degrees of freedom. The corresponding
degrees of freedom $df_{i}$ for $S_{i}^{2}$ can be determined by
Satterthwaite approach (Satterthwaite, 1946):
\begin{equation}
df_{i}=S_{i}^{4}/[S_{i0}^{4}/(n_{i0}^{2}(n_{i0}-1))+S_{i1}^{4}/(n_{i1}^{2}(n_{i1}-1))].
\end{equation}

The simulation model deviates from the model assumptions described
in Section \ref{sec:Statistical-methods}, due to the introduction
of the random variable $V_{i}$. First of all, the marginal distribution
of $Y_{i}$ is no longer normal, although the conditional distribution
of $Y_{i}$ given $V_{i}$ is normally distributed with mean $\theta$
and variance $\exp(2V_{i})[\xi_{0}^{2}/n_{i0}+\xi_{1}^{2}/n_{i1}]$.
Secondly, the variance of $Y_{i}$ given $S_{i}^{2}$ is unequal to
$S_{i}^{2}$, since the conditional distributions of $Y_{i}$ and
$S_{i}^{2}$ given $V_{i}$ are independent (see van den Heuvel et
al. (2020)). Finally, the marginal distribution of $S_{i}^{2}$ is
not directly related to a chi-square distribution. Only the conditional
distribution of $S_{i}^{2}$ given $V_{i}$ is approximately chi-square
distributed using Satterthwaite approach (Satterthwaite, 1946), i.e.,
$df_{i}S_{i}^{2}/[\exp\{V_{i}\}(\xi_{0}^{2}/n_{i0}+\xi_{1}^{2}/n_{i1})]$
is approximately chi-square distributed with $df_{i}$ degrees of
freedom conditioned on $V_{i}$. And it becomes exactly chi-square
distributed with $n_{i0}+n_{i1}-2$ degrees of freedom conditioned
on $V_{i}$, when both $\xi_{0}=\xi_{1}$ and $n_{i0}=n_{i1}$ hold.
The marginal distribution of $S_{i}^{2}$ is less traceable. Thus
all proposed estimation methods with their accompanied confidence
intervals for $\theta$ in Section \ref{sec:Statistical-methods}
are at best approximate methods for the data of our simulated meta-analysis.
We believe that none of the approaches has an obvious direct advantage
over any of the other methods.

The settings of the parameters are chosen such that the simulation
corresponds approximately with a meta-analysis of clinical trials
on for instance hypertension treatment (for systolic blood pressure).
Parameter settings used to generate the aggregated data $(Y_{i},S_{i},df_{i})$
from the individual participant data are $m\in\{10,20.30\}$, $\lambda=100$,
$a_{0}=b_{0}=1$, $p=0.5$, $\mu=160$, $\theta=-2$, $\xi_{0}^{2}=\xi_{1}^{2}=100$.
We will run several combinations of the remaining parameters $\sigma_{0}^{2}$,
$\sigma_{1}^{2}$, $\sigma_{2}^{2}$, $\rho_{M}$ and $\rho_{V}$
of the IPD model:
\begin{enumerate}
\item \textbf{\uline{Setting 1:}} Homogeneous study effects and no random
heteroscedastic residuals: $\sigma_{0}^{2}=0$, $\sigma_{1}^{2}=0$,
$\sigma_{2}^{2}=0$, $\rho_{M}=0$ and $\rho_{V}=0$,
\item \textbf{\uline{Setting 2:}} Heterogeneous study effects and no
random heteroscedastic residuals: $\sigma_{0}^{2}=2$, $\sigma_{1}^{2}=3$,
$\sigma_{2}^{2}=0$, $\rho_{M}=0.7$, and $\rho_{V}=0$,
\item \textbf{\uline{Setting 3:}} Heterogeneous study effects and random
heteroscedastic residuals without correlation: $\sigma_{0}^{2}=2$,
$\sigma_{1}^{2}=3$, $\sigma_{2}^{2}=1$, $\rho_{M}=0.7$, and $\rho_{V}=0$,
\item \textbf{\uline{Setting 4:}} Heterogeneous study effects and random
heteroscedastic residuals with low correlation: $\sigma_{0}^{2}=2$,
$\sigma_{1}^{2}=3$, $\sigma_{2}^{2}=1$, $\rho_{M}=0.7$, and $\rho_{V}=0$.3,
\item \textbf{\uline{Setting 5:}} Heterogeneous study effects and random
heteroscedastic residuals with medium correlation: $\sigma_{0}^{2}=2$,
$\sigma_{1}^{2}=3$, $\sigma_{2}^{2}=1$, $\rho_{M}=0.7$, and $\rho_{V}=0.5$,
\item \textbf{\uline{Setting 6:}} Heterogeneous study effects and random
heteroscedastic residuals with high correlation: $\sigma_{0}^{2}=2$,
$\sigma_{1}^{2}=3$, $\sigma_{2}^{2}=1$, $\rho_{M}=0.7$, and $\rho_{V}=0$.7.
\end{enumerate}

\section{Results}

Tables \ref{tab:Bias} and \ref{tab:MSE} present the bias and the
MSE of the three different estimation methods (DL, HT, and BD), respectively.
Note that the Noma-Bartlett and Guolo-Skovgaard confidence intervals
make use of the maximum likelihood estimators of Hardy and Thompson.

For the settings without heteroscedasticity (settings 1 and 2) the
biases of DL, HT, and BD are all similar, irrespective of sample size
biases remain within 1.2\% of the true effect size ($\theta=-2$)
for the homogeneous study effect sizes. In the presence of uncorrelated
heterogeneous study effect sizes and random heteroscedasticity (setting
3), again all biases are very close to zero for all three sample sizes.
However, in case of correlated heterogeneous study effect sizes and
random heteroscedasticity (settings 4 through 6), only BD seems to
have small biases for all sample sizes and it is never larger than
0.9\% of the true effect size. The biases of DL and HT are away from
zero, in particular when the correlation between the heterogeneous
study effect sizes are strongly correlated to the random heteroscedasticity.
The sample size does not seem to affect this. The bias can then reach
a level of 5\% of the true effect size. For $m=10$, BD and DL are
similar and very close to zero. The performance of DL seems to be
the worst for $m=20$, with a bias that could reach 1.5\%. Unfortunateley,
HT seem to provide a small negative bias for all three study sizes
that can reach more than 5\%.
\begin{table}[h]
\caption{\label{tab:Bias}Bias of the estimation methods under different simulation
settings and for $\theta=-2$.}

\centering

\begin{tabular}{|c||c|c|c||c|c|c||c|c|c|}
\hline
\multirow{2}{*}{Setting} & \multicolumn{3}{c||}{$m=10$} & \multicolumn{3}{c||}{$m=20$} & \multicolumn{3}{c|}{$m=30$}\tabularnewline
\cline{2-10}
 & DL & HT & BD & DL & HT & BD & DL & HT & BD\tabularnewline
\hline
\hline
1 & -0.024 & -0.024 & -0.024 & -0.021 & -0.021 & -0.020 & -0.010 & -0.010 & -0.011\tabularnewline
\hline
2 & -0.006 & -0.008 & -0.007 & -0.013 & -0.015 & -0.012 & -0.004 & -0.004 & -0.004\tabularnewline
\hline
3 & 0.007 & 0.010 & -0.010 & -0.001 & 0.001 & -0.018 & -0.001 & -0.001 & -0.006\tabularnewline
\hline
4 & -0.034 & -0.036 & -0.007 & -0.043 & -0.044 & -0.017 & -0.045 & -0.046 & -0.007\tabularnewline
\hline
5 & -0.062 & -0.068 & -0.006 & -0.073 & -0.075 & -0.016 & -0.075 & -0.077 & -0.007\tabularnewline
\hline
6 & -0.091 & -0.098 & -0.005 & -0.103 & -0.107 & -0.015 & -0.106 & -0.108 & -0.006\tabularnewline
\hline
\end{tabular}
\end{table}

The performance of MSE for the three estimation methods is very consistent
across all settings. For all methods, the MSE increases with settings,
which is expected due to the increased variability. Setting 1 has
no study heterogeneity and no random heteroscedasticity, and thus
the smallest variability across meta-analysis studies. Setting 2 has
heterogeneous study effect sizes but no heteroscedasticity yet. Then
for settings three to six, the residual variance increases due to
the random heteroscedasticity and an increased positive correlation
$\rho_{V}$, while the heterogeneity in study effect sizes remains
constant (although the correlation seem to have little effect). When
no random heteroscedasticity is present, the MSE of the three estimation
approaches DL, HT, and BD are almost identical. However, when heteroscedasticity
is present, the MSE of BD is larger than the MSE of DL and HT. The
MSE of DL and HT seem to be identical across all settings and sample
sizes. It seems that the random heteroscedasticity does hardly affect
the MSE of DL and HT, since it is at the same level as setting 2 which
had no random heteroscedasticity, but BD is strongly affected.
\begin{table}[h]
\caption{\label{tab:MSE}MSE of the estimation methods under different simulation
settings and for $\theta=-2$.}

\centering

\begin{tabular}{|c||c|c|c||c|c|c||c|c|c|}
\hline
\multirow{2}{*}{Setting} & \multicolumn{3}{c||}{$m=10$} & \multicolumn{3}{c||}{$m=20$} & \multicolumn{3}{c|}{$m=30$}\tabularnewline
\cline{2-10}
 & DL & HT & BD & DL & HT & BD & DL & HT & BD\tabularnewline
\hline
\hline
1 & 0.239 & 0.238 & 0.232 & 0.105 & 0.104 & 0.102 & 0.069 & 0.069 & 0.068\tabularnewline
\hline
2 & 0.435 & 0.435 & 0.429 & 0.197 & 0.197 & 0.196 & 0.128 & 0.128 & 0.127\tabularnewline
\hline
3 & 0.434 & 0.438 & 0.588 & 0.202 & 0.203 & 0.280 & 0.127 & 0.127 & 0.181\tabularnewline
\hline
4 & 0.431 & 0.435 & 0.605 & 0.202 & 0.204 & 0.276 & 0.126 & 0.127 & 0.182\tabularnewline
\hline
5 & 0.432 & 0.436 & 0.610 & 0.204 & 0.206 & 0.271 & 0.128 & 0.129 & 0.182\tabularnewline
\hline
6 & 0.436 & 0.441 & 0.617 & 0.208 & 0.210 & 0.265 & 0.132 & 0.132 & 0.183\tabularnewline
\hline
\end{tabular}
\end{table}

Figure 1 presents the coverage probabilities for the five methods
on calculation of 95\% confidence intervals on the main parameter
$\theta$ for the six different simulation settings. For the (unrealistic)
case of homogeneous effect sizes without random heteroscedasticity
(setting 1) the methods show above nominal coverage probabilities,
although the HT method seems closer to nominal than the others. When
the number of studies is $m=30$ all coverages are very close to 97\%.
The DL, BD, NB, and GS method seem to decrease to this coverage when
study sizes increase from $m=10$ to $m=30$, while HT show a small
increase to this coverage. In case random heteroscedasticity is introduced,
the DL and HT method seem to underperform and provide liberal coverage
probabilities, while DB, NB, and GS seem to provide coverages (very)
close to the nominal 95\% coverage, although the GS method seem to
be slightly, but consistently, conservative at $m=10$ studies. For
the heterogeneous effect sizes with no random heteroscedasticity (setting
2), all methods seem close to the nominal coverage of 95\%, in particular
when the number of studies is $m=20$ or larger.
\begin{figure}[H]
\raggedright{}\hfill{}\subfloat[Homogeneous effect sizes with no random heteroscedasticity (setting
1).]{\begin{centering}
\hfill{}
\par\end{centering}
\centering{}\includegraphics[scale=0.31]{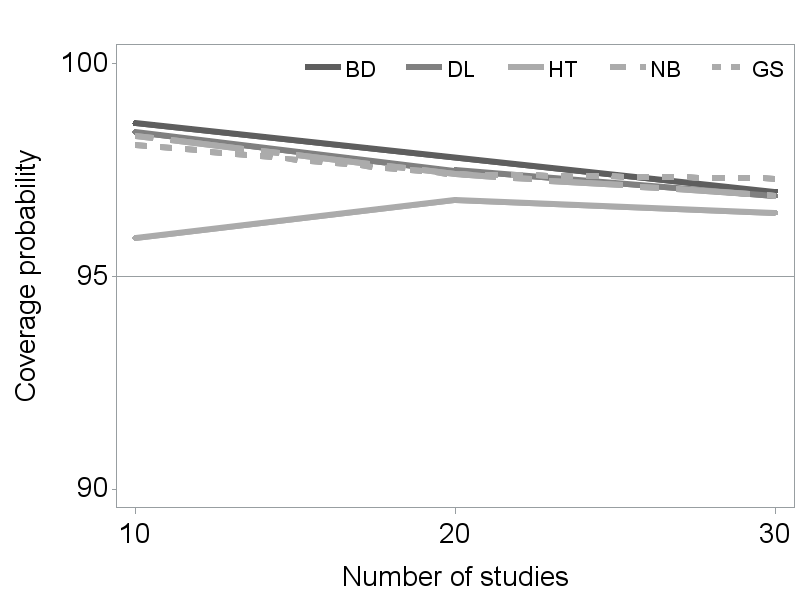}}\hfill{}\subfloat[Heterogeneous effect sizes with no random heteroscedasticity (setting
2).]{\begin{centering}
\hfill{}
\par\end{centering}
\centering{}\includegraphics[scale=0.31]{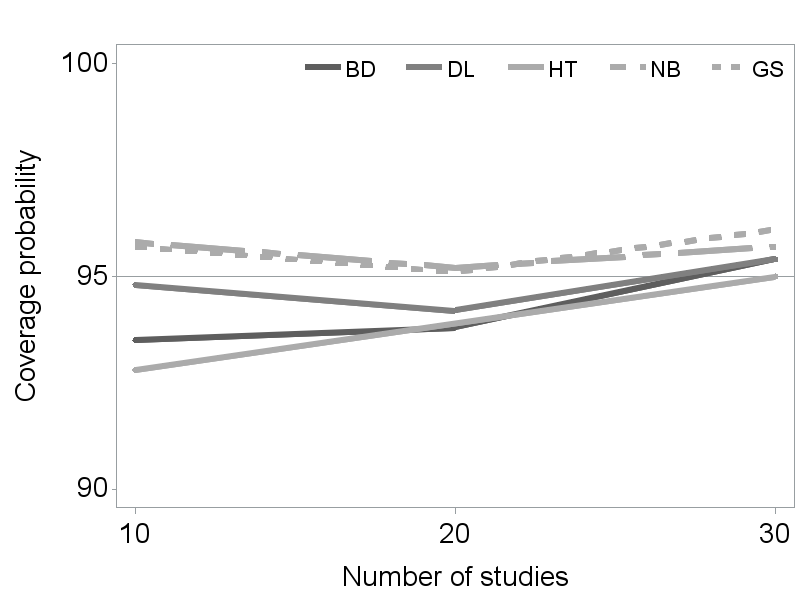}}
\end{figure}

\begin{figure}[H]
\raggedright{}\hfill{}\subfloat[Heterogeneous and heteroscedastic effect sizes with $\rho_{02}=\rho_{12}=0$
(setting 3).]{\begin{centering}
\hfill{}
\par\end{centering}
\centering{}\includegraphics[scale=0.31]{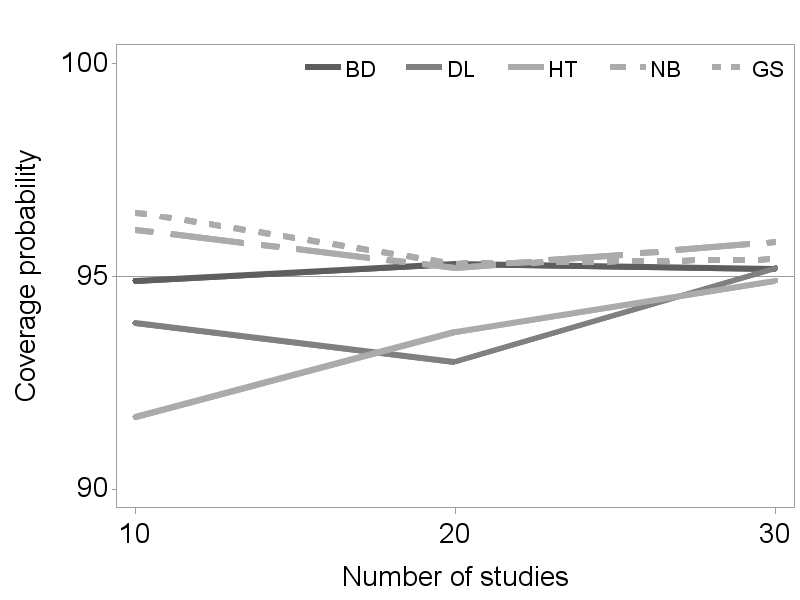}}\hfill{}\subfloat[Heterogeneous and heteroscedastic effect sizes with $\rho_{02}=\rho_{12}=0.3$
(setting 4).]{\begin{centering}
\hfill{}
\par\end{centering}
\centering{}\includegraphics[scale=0.31]{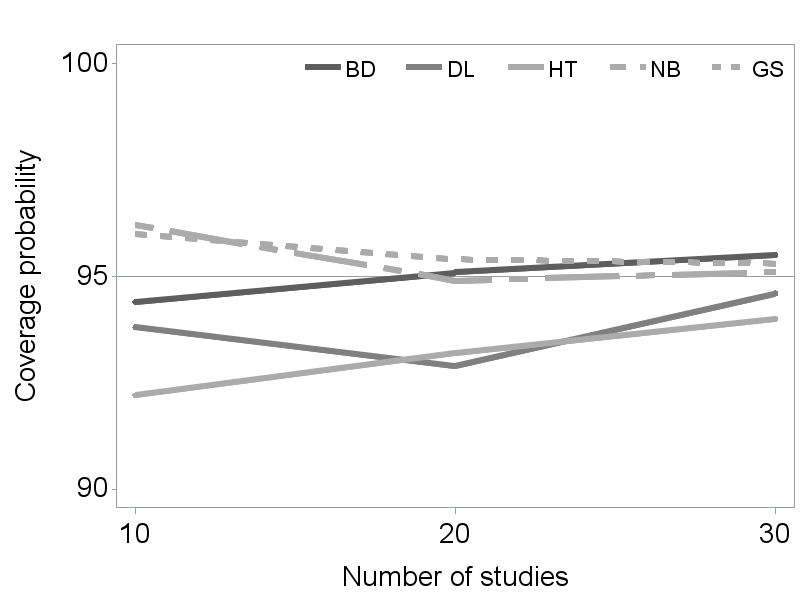}}
\end{figure}

\begin{figure}[H]
\begin{raggedright}
\hfill{}\subfloat[Heterogeneous and heteroscedastic effect sizes with $\rho_{02}=\rho_{12}=0.5$
(setting 5).]{\begin{centering}
\hfill{}
\par\end{centering}
\centering{}\includegraphics[scale=0.31]{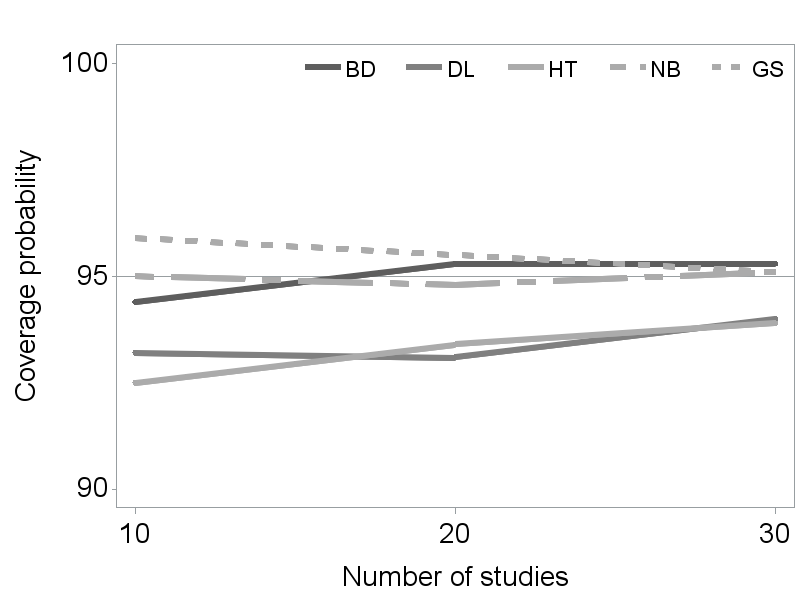}}\hfill{}\subfloat[Heterogeneous and heteroscedastic effect sizes with $\rho_{02}=\rho_{12}=0.7$
(setting 6).]{\begin{centering}
\hfill{}
\par\end{centering}
\centering{}\includegraphics[scale=0.31]{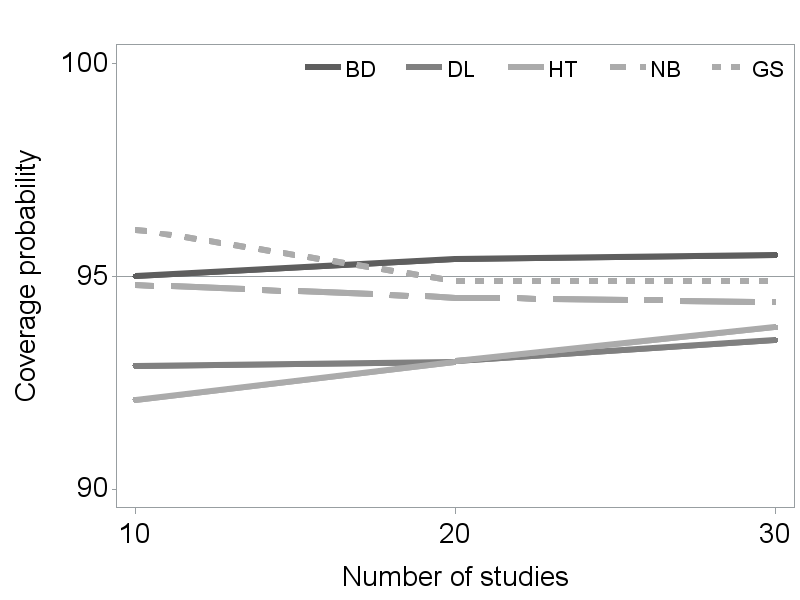}}
\par\end{raggedright}
\caption{Empirical coverage percentages of 95\% confidence intervals of five
methods for the overall effect size under different settings and study
sizes.}
\end{figure}

To complete the comparison, we also compared the estimates of the
between study variance $\tau^{2}$ for the three estimation methods
DL, HT, and BD. For the first setting the variance $\mathsf{VAR}(U_{i})=\mathsf{VAR}(U_{i0}-U_{i1})$
is $\tau^{2}=0$ and in the remaining settings this variance is $\tau^{2}=\sigma_{0}^{2}+\sigma_{1}^{2}-2\rho_{M}\sigma_{0}\sigma_{1}=2+3-2\times0.7\times\sqrt{2}\times\sqrt{3}\approx1.5707$.
However, in case of heteroscedasticity, the correlation between the
heterogeneous study effect sizes $U_{i}$ and the random heteroscedasticity
$V_{i}$ may affect the estimation of the between study variance,
but we expect it to be still close to 1.5707. The results of the estimates
are presented in the following table.
\begin{table}[h]
\caption{\label{tab:Tau2}Between study variances of the three estimation methods
under different simulation settings.}

\centering

\begin{tabular}{|c||c|c|c||c|c|c||c|c|c|}
\hline
\multirow{2}{*}{Setting} & \multicolumn{3}{c||}{$m=10$} & \multicolumn{3}{c||}{$m=20$} & \multicolumn{3}{c|}{$m=30$}\tabularnewline
\cline{2-10}
 & DL & HT & BD & DL & HT & BD & DL & HT & BD\tabularnewline
\hline
\hline
1 & 0.436 & 0.281 & 0.238 & 0.296 & 0.207 & 0.172 & 0.240 & 0.177 & 0.146\tabularnewline
\hline
2 & 1.779 & 1.404 & 1.306 & 1.601 & 1.410 & 1.330 & 1.589 & 1.468 & 1.393\tabularnewline
\hline
3 & 1.728 & 1.320 & 1.661 & 1.560 & 1.380 & 1.629 & 1.569 & 1.449 & 1.615\tabularnewline
\hline
4 & 1.725 & 1.325 & 1.683 & 1.564 & 1.379 & 1.625 & 1.575 & 1.445 & 1.581\tabularnewline
\hline
5 & 1.723 & 1.332 & 1.674 & 1.565 & 1.380 & 1.629 & 1.576 & 1.441 & 1.556\tabularnewline
\hline
6 & 1.722 & 1.333 & 1.675 & 1.563 & 1.376 & 1.614 & 1.563 & 1.439 & 1.549\tabularnewline
\hline
\end{tabular}
\end{table}

Without heterogeneity and random heteroscedasticity, all methods are
biased, but the BD method is closest to the truth and the bias reduces
with sample size. In case of heterogeneity, but without random heteroscedasticity,
the DL approach is closest to the true value when sample sizes $m$
are 20 or larger. DL seems to overestimate the variance, while HT
and BD underestimates the variance. This latter observation is well
known characteristic of maximum likelihood estimation for variance
components. In case of heterogeneity and random heteroscedasticity,
the BD and DL method are closer to the truth than the HT method. The
BD method is better than DL when sample sizes are small, while DL
is slightly better than BD when sample sizes are larger. The HT method
seems to be biased in all settings.

\section{Discussion}

The purpose of this article was to introduce a joint analysis of the
study effect sizes and its estimated standard error for aggregated
data meta-analyses. A combination of a normal and chi-square distribution
was used to describe the distribution of the observed bivariate statistics.
The performance of this bivariate distribution was compared to that
of the DerSimonian-Laird method and three likelihood-based methods.
The likelihood-based methods assumed that the residual variance of
the study effect size is equal to the squared standard error. We studied
the profile likelihood approach of Hardy and Thompson, the Bartlett-corrected
likelihood ratio, and the Skovgaard corrected likelihood ratio. A
simulation study with different scenarios was carried out using different
numbers of studies and different correlation structures between the
study effect sizes and its standard error. The simulation settings
explicitly studied (random) heteroscedasticity of the true residual
variance of the study effect sizes, because we believe that heteroscedasticity
is common in practice. None of the five studied approaches are equipped
to deal with this heteroscedasticity explicitly.

Differences between the methods for estimation of the pooled effect
size with its accompanied confidence interval were relatively small,
but some differences were observed. When heteroscedasticity is introduced,
the DerSimonian-Laird and Hardy-Thompson approach show a small bias
in the pooled effect size. This bias most likely caused a liberal
coverage probability, a conclusion already established in literature
(Norma, 2011; Guolo,2012). In case we apply the Hartung-Knapp-Sidak-Jonkman
standard error estimate for the DerSimonian-Laird method, the coverage
improves for the homogeneous and homoscedastic setting, but it remains
similar to the DL results for all other settings (data not shown).
Since the corrected likelihood approaches use a finite sample approximation
of the distribution of the Hardy-Thompson estimator, these corrected
approaches provide the same bias as the Hardy-Thompson method, but
they do improve the coverage probability. The Bartlett-type and the
Skovgaard corrected likelihood ratio methods have comparable results,
and are slightly conservative when the number of studies is small,
but for larger study sizes they provide nominal coverage probabilities.
These conclusions have been established earlier too (Norma, 2011;
Veroniki et al., 2019). More generally, all methods provide nominal
coverages as the number of studies increases. Our bivariate approach
provided similar and consistent results in all performance measures
under heterogeneity and random heteroscedasticity, with coverage probabilities
close to nominal for all sample sizes. The coverage is very similar
or better than the two finite sample size corrected likelihood approaches
and outperforms DerSimonian-Laird and Hardy-Thompson approaches. The
disadvantage of our approach is the need for a degrees of freedom,
but the analysis is straightforward and based on first-order asymptotics
that do not need a finite sample correction. It also performs well
when studies are heterogeneous in both the study effect sizes and
their standard errors.

\section*{Acknowledgements}

This research was funded by grant number 023.005.087 from the Netherlands
Organization for Scientific Research.

\section*{Conflict of interest}

The authors have declared no conflict of interest.

\section*{\newpage Appendix: code used for implementing the Bivariate distribution
method}

The following programming codes in proc NLMIXED assume that there
exists a data set ``Effect\_Sizes'' with different columns and rows.
The rows represent studies which are listed in column ``Study''.
For each study we have two separate rows: one row for the effect size
$Y_{i}$ and a second row for the variance $S_{i}^{2}$. The effect
size $Y_{i}$ and variance $S_{i}^{2}$ are below eachother in the
same column called ``Outcome'' and to identify these different responses
we have a column ``Response'' with levels ``effect size'' and
``variance''. Finally, there is a column with the degrees of freedom
for each study. Table \ref{tab:Data-overview-1} shows schematically
how the data is organized.

\begin{table}[h]
\caption{\label{tab:Data-overview-1}Schematic overview of how the data of
a meta-analysis should be organized to execute our bivariate distribution
approach.}

\begin{tabular}{|c|c|c|c|}
\hline
Study & Response & Outcome & Degrees\tabularnewline
\hline
\hline
1 & effect size & $Y_{1}$ & $df_{1}$\tabularnewline
\hline
1 & variance & $S_{1}^{2}$ & $df_{1}$\tabularnewline
\hline
2 & effect size & $Y_{2}$ & $df_{2}$\tabularnewline
\hline
2 & variance & $S_{2}^{2}$ & $df_{2}$\tabularnewline
\hline
$\vdots$ & $\vdots$ & $\vdots$ & $\vdots$\tabularnewline
\hline
$\vdots$ & $\vdots$ & $\vdots$ & $\vdots$\tabularnewline
\hline
$m$ & effect size & $Y_{m}$ & $df_{m}$\tabularnewline
\hline
$m$ & variance & $S_{m}^{2}$ & $df_{m}$\tabularnewline
\hline
\end{tabular}
\end{table}

PROC NLMIXED DATA = Effect\_Sizes QPOINTS=10 MAXITER=100 TECH=NEWRAP;

PARMS THETA = 0 LNSTAU = 0 SD = 10;

MU = THETA + U;

TAU2 = EXP(2{*}LNSTAU);

VAR\_I = (SD{*}{*}2)/Degrees;

IF Response = ``effect size'' THEN DENS = -0.5{*}LOG(2{*}3.14159)-0.5{*}LOG(VAR\_I)-0.5{*}((Outcome
- MU){*}{*}2)/VAR\_I;

ELSE IF Response = ``variance'' THEN DENS = -(Degrees/2){*}LOG(2)-LGAMMA(Degrees/2)+((Degrees/2)-1){*}LOG(Degrees{*}Outcome/VAR\_I)-0.5{*}Degrees{*}Outcome/VAR\_I;

MODEL Outcome\textasciitilde GENERAL(DENS);

RANDOM U \textasciitilde{} NORMAL (0, TAU2) SUBJECT = Study;

RUN;QUIT;

\newpage{}

\end{document}